\documentclass{ws-ijmpe}

\usepackage{bm}% bold math

\newcommand{\be}{\begin{equation}}
\newcommand{\ee}{\end{equation}}
\newcommand{\ben}{\begin{eqnarray}}
\newcommand{\een}{\end{eqnarray}}
\newcommand{\ba}{\begin{array}}
\newcommand{\ea}{\end{array}}
\newcommand{\bc}{\begin{center}}
\newcommand{\ec}{\end{center}}
\newcommand{\bml}{\begin{mathletters}}
\newcommand{\eml}{\end{mathletters}}

\begin{document}

\markboth{M. Zalewski, W. Satu{\l}a}{Terminating states as a unique laboratory
for testing nuclear energy density functional}

\catchline{}{}{}{}{}
%
%%%%%%%%%%%%%%%%%%%%%%%%%%%%%%%%%%%%%%%%%%%%%%%%%%%%%%%%%%%%%%%%%%%%

\author{\footnotesize M. ZALEWSKI\footnote{zalewiak@fuw.edu.pl}}
\address{Institute of Theoretical Physics, University of Warsaw, \\
ul. Ho\.za 69, 00-681 Warsaw, Poland\\
}

\author{W. SATU{\L}A\footnote{satula@fuw.edu.pl}}
\address{Institute of Theoretical Physics, University of Warsaw, \\
ul. Ho\.za 69, 00-681 Warsaw, Poland,\\
%KTH (Royal Institute of Technology)\\
%AlbaNova University Center, 106 91 Stockholm, Sweden\\
}

\title{TERMINATING STATES AS A UNIQUE LABORATORY FOR
TESTING NUCLEAR ENERGY DENSITY FUNCTIONAL
}

\maketitle
\begin{history}
\received{(received date )}
\revised{(revised date )}
%\accepted{(Day Month Year)}
%\comby{(xxxxxxxxxx)}
\end{history}

\begin{abstract}
Systematic calculations of favored signature maximum-spin
$I_{max}$ and unfavored signature $I_{max}-1$
terminating states for $[f_{7/2}^n]$ and   $[d_{3/2}^{-1} f_{7/2}^{n+1}]$
configurations ($n$ denotes number of valence particles)
in $A$$\sim$44 mass region are presented.
Following the result of Zdu\'nczuk {\it et al.\/},
Phys. Rev. {\bf C71} (2005) 024305 the calculations
are performed using Skyrme energy density functional
with empirical Landau parameters and slightly reduced spin-orbit
strength. The aim is to identify and phenomenologically restore rotational
symmetry broken by the Skyrme-Hartree-Fock solutions. In particular,
it is shown that correlation energy due to symmetry restoration
is absolutely crucial in order to reproduce
energy splitting $E(I_{max}) - E(I_{max}-1)$ in
$[f_{7/2}^n]$ configurations
but is relatively less important for $[d_{3/2}^{-1}f_{7/2}^{n+1}]$
configurations.
\end{abstract}

\section{Introduction}

 Structure of heavy nuclei may be described using two major
theoretical approaches: the nuclear shell-model~\cite{[Cau05]} (SM) and the
nuclear density functional theory~\cite{[Ben03]} (DFT). The main
advantage of SM is proper treatment of many-body correlations among limited
number of valence particles. The resulting wave functions
are eigenstates to symmetry invariants of the effective Hamiltonian.
In the language of
DFT, nucleus is treated as $A$-body composite object described by one-body
densities and currents. According to the Hohenberg-Kohn-Sham (HKS)
theorem~\cite{[Koh98x]}, this method is formally exact. However, since
the HKS theorem does not provide any method of constructing
such an exact functional standard procedures employed in nuclear physics
attempt to construct the functional in a systematic way guided
by basic symmetry requirements~\cite{[Per04]}.
A particular case of the nuclear EDF is the functional inspired
by well studied Skyrme model~\cite{[Ben03],[Sky56x],[Vau72]}
(S-EDF) which, in the isoscalar-isovector $t=0,1$ representation, takes the
following form:
\be\label{efun} {\cal
E}^{Skyrme}=\sum_{t=0,1}\int
  d^3{\boldsymbol r}
\left[ {\mathcal H}_t^{(TE)} ({\boldsymbol r}) +
       {\mathcal H}_t^{(TO)} ({\boldsymbol r}) \right],
\ee
where
\be\label{teven} {\mathcal
H}_{t}^{(TE)} ({\boldsymbol r}) = C_{t}^{\rho} \rho_t^2
+ C_{t}^{\Delta\rho} \rho_t\Delta\rho_t + C_t^\tau \rho_t\tau_t + C_t^J
{J}_t^2 + C_t^{\nabla J} \rho_t
{\boldsymbol \nabla}\cdot {\boldsymbol J}_t,
\ee
\be\label{todd}
{\mathcal H}_{t}^{(TO)}
({\boldsymbol r}) = C_{t}^{s}
\boldsymbol{s}_t^2 + C_{t}^{\Delta s} \boldsymbol{s}_t\Delta \boldsymbol{s}_t
+ C_t^T \boldsymbol{s}_t\cdot \boldsymbol{T}_t + C_t^j {\boldsymbol j}_t^2 +
C_t^{\nabla j} \boldsymbol{s}_t  \cdot ( {\boldsymbol \nabla}\times
{\boldsymbol j}_t).
\ee
The functional ${\mathcal H}$ is uniquely
expressed as a bilinear form of time-even ({\sc te}) $\rho, \tau,
{J}$  and time-odd ({\sc to}) $\boldsymbol{s},
\boldsymbol{T}, \boldsymbol{j}$ local densities, currents, and by their
derivatives. Exact expressions linking
auxiliary Skyrme parameters $x_i, t_i, \, i=0,1,2,3$ and $W, \alpha$
with coupling constants $C$ can be found, for example,
in Ref.~\cite{[Ben03]}.

The coupling constants of the S-EDF (or Skyrme force (SF)) are
adjusted to global properties of nuclear matter and to
selected experimental observables in order to account for basic
nuclear structure properties. Since there exist a multitude of
observables and no real consensus about which of
these are to be selected in a unique manner, there exist a multitude of
different SF parameterizations. One of the basic problems in adjusting the
force parameters is related to the fact that the single-particle
(SP) states, that
are so crucial for high accuracy calculations, are in general coupled to
collective motion and therefore difficult to determine. In this context, any
dataset representing SP motion
is an invaluable source of information that can be used
for a rigorous test and subsequent fine-tuning of the parameters.

Our long standing experience tells us that
terminating states which are maximum-spin states within given SP
configuration are one of the best examples of unperturbed  SP
motion~\cite{[Afa99a],[Sat05]}.
This conclusion was recently confirmed by a comparative study between
state of the art SM  and Skyrme-Hartree-Fock calculation~\cite{[Sto06]}
performed for maximum spin $I_{max}$ states terminating
within  $f_{7/2}^n$ and $d_{3/2}^{-1}f_{7/2}^{n+1}$ configurations
in $N$$>$$Z$, $A$$\sim$44 nuclei. Indeed, for these cases the
energy difference
\be\label{deltae}
\Delta E =
 E[d_{3/2}^{-1}f_{7/2}^{n+1}]_{I_{max}} - E[f_{7/2}^n]_{I_{max}} \, ,
\ee
calculated using  Skyrme-Hartree-Fock (SHF) approach with modified
spin-fields and spin-orbit (SO) strength~\cite{[Zdu05y]}
follows very closely the results of SM calculations.
This study revealed  simultaneously that in $N$=$Z$ nuclei the
extreme SP picture breaks down due to spontaneous breaking
of isobaric symmetry.

\smallskip

The aim of this work is to explore the
structural simplicity of maximum-spin
$I_{max}$ (favored signature) as well as for $I_{max}-1$ (unfavored signature)
terminating states in order to further constrain parameters
of the S-LEDF as well as in order to identify, quantitatively
evaluate and subsequently restore in a  phenomenological
way broken symmetries inherently obscuring the SHF treatment.
In particular, we will show that the energy difference
of Eq.~(\ref{deltae}) between $I_{max}$ configurations
$p_{1/2}^{-1}d_{5/2}^{n+1}$ and $d_{5/2}^{n}$
in $A$$\sim$16 mass region and between  $I_{max}$ configurations
$f_{5/2}^{-1}g_{9/2}^{n+1}$ and $g_{9/2}^{n}$
in $A$$\sim$80 mass region follows closely the trend found
in $A$$\sim$44 mass region in Refs.~\cite{[Zdu05y]} with respect to
the isoscalar effective mass scaled isoscalar strength of
SO term and that this result nicely correlates with
simple Nilsson model predictions concerning $N$=$Z$=8 and
$N$=$Z$=40 magic gaps.
In the second part we will demonstrate that
unfavored signature $I_{max}-1$
terminating states in $A$$\sim$44 mass region manifestly violate
rotational invariance. However, unlike in most other cases
where the spontaneous breaking of rotational symmetry (SSB)
leads to the occurrence of deformation what allows to incorporate
substantial fraction of many-body correlations into a
single deformed Slater determinant, see Ref.~\cite{[Sat05]}
and refs. quoted therein, the effect discussed here
occurs at quasi-spherical shape. It appears that
there is a fundamental difference between these two situations.
In deformed nuclei the SSB mechanism works
{\it constructively\/}. The violated symmetry can be
subsequently approximately restored within independent particle
model based on cranking approximation what leads naturally to
emergence of collective rotational states.
In the case of $I_{max}-1$ terminating states
the SSB mechanism works {\it destructively\/}
and one need to go beyond mean-field and perform configuration mixing
calculations to restore the symmetry. However, since the number
of participating (dominant) configurations is very limited
such configuration mixing can be performed
analytically, as shown in detail in Sect.~\ref{i-1}.

\section{Isoscalar spin-orbit strength}

In Refs.~\cite{[Zdu05y]} it was shown that the set of terminating
states in $N$$\ne$$Z$, $A$$\sim$44 mass region provides reliable constraints
on time-odd fields and the strength of isoscalar SO
interaction of the S-EDF. In particular, it was demonstrated that
constraining coupling constants of time odd spin-fields to the
empirical spin-isospin Landau parameters leads to unification of predictions
for $\Delta E$ of Eq.~(\ref{deltae}) for such different
Skyrme parameterizations like SLy4 and SLy5~\cite{[Cha97]}, SkO~\cite{[Rei99]},
SIII~\cite{[Bei75]}, and SkXc~\cite{[Bro98]} with mean deviation between
calculated and experimental values
$\overline{\Delta E}$$\sim$500\,keV  i.e. at the level of $\sim$10\%.
Let us stress that our high-spin estimate of Landau parameters is
fairly consistent with the data extracted based on completely different
experimental input like giant resonances, beta decays, or
moments of inertia~\cite{[Ost92],[Ben02a]}.

The remaining mean discrepancy
$\overline{\Delta E}\equiv \overline{\Delta E}_{exp} -
\overline{\Delta E}_{th}$ between theory and
experiment can be further reduced by slightly reducing
SO strength. Indeed, the quantity $\Delta E$ of Eq.~(\ref{deltae})
is governed by the size of $N=Z=20$ magic gap $\Delta e_{20}$ which, in turn,
strongly depends on the SO strength. It can be nicely demonstrated
within the spherical Nilsson Hamiltonian~\cite{[Nil55]} where:
\be
\Delta e_{20}= \hbar\omega_0(1-6\kappa-2\kappa\mu).
\ee
For light nuclei $\mu \sim 0$ i.e. flat bottom and surface effect do
not play important role. The width of the potential well
$\hbar \omega_0$ determines global energy scale in low energy
nuclear physics. It is rather well constrained by data and even small
variations can spoil in general good agreement between theory and
experiment, especially in heavy nuclei. Hence, the SO term
$\kappa$ plays indeed a dominant role in magnitude of the magic gap
$\Delta e_{20}$.

%%%%%%%%%%%%%%%%%%%%%%%%%%%%%%%%%%%%%%%%%%%%%%%%%%%%%%%%%%%%%%%%%%%%%%%%%%%%
\begin{figure}[htb]\label{W0}
\centering
\includegraphics[width=8cm,clip]{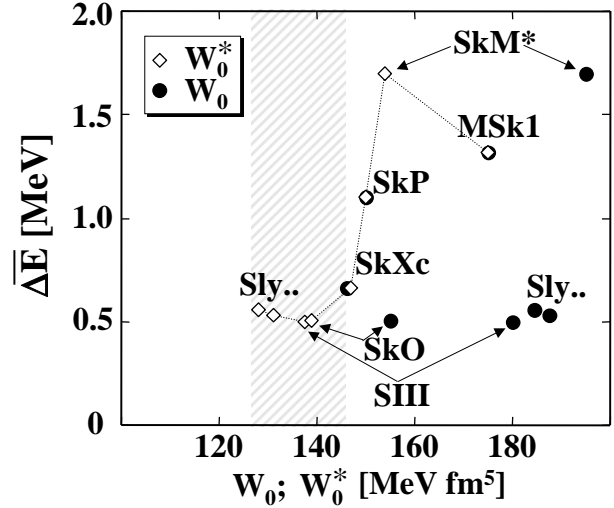}
\caption[]{Mean value $\overline{\Delta E}$ as a function of the isoscalar
strength $W_0^*$ (filled
diamonds) and $W_0$ (open circles) for different parameterizations of
the SF. See text for further discussion.
Taken from Ref.~\protect\cite{[Zdu05y]}.}
\end{figure}
%%%%%%%%%%%%%%%%%%%%%%%%%%%%%%%%%%%%%%%%%%%%%%%%%%%%%%%%%%%%%%%%%%%%%%%%%%%%%

In Refs.~\cite{[Zdu05y]} it was shown, that $\overline{\Delta E}$
in $A$$\sim$44 nicely correlates with isoscalar SO strength.
However, due to non-local momentum dependent terms,
such a  correlation cannot be done at the level of
bare SO strength $W_0$, see black dots in Fig.~\ref{W0}, but must be
performed at the level of the  so called asymptotically equivalent
representation:
\be
     \tilde{\phi_i}(\vec{r})=\sqrt{\frac{m}{m^*(\vec{r})}} \phi_i(\vec{r}).
\ee
In this representation free particles in the infinity $r\rightarrow\infty$
acquire not effective $m*$ but bare mass
$m$ while the SO potential takes the form:
\be
        V_{LS}(q,r)  \approx
        \frac{m^*(\vec{r})}{m}
        \left\{W_0 \frac{1}{r}\rho_0'(r) \pm
        W_1 \frac{1}{r}\rho_1'(r)\right\}\vec{l}\vec{s}\, .
\ee
Note, that the true SO strength felt by a nucleon is $W_0^* \equiv
\frac{m^*}{m} W_0$ and not $W_0$. The correlation between
the effective-mass-scaled isoscalar strength is  $W_0^*$
and $\overline{\Delta E}$ is now evident, see diamonds in Fig.\ref{W0}.
All forces giving similar $\overline{\Delta E}$
including SLy4, Sly5, SIII, SkO and SkXc have also similar
$W_0^*$ ($\sim 135 \pm 10 \, \mathrm{MeV \, fm^5}$).
For SkP~\cite{[Dob84]}, MSk1~\cite{[Ton00]} or
SkM*~\cite{[Bar82]} which give unacceptably large
$\overline{\Delta E}$, $W_0^*$ is also considerably larger.

%%%%%%%%%%%%%%%%%%%%%%%%%%%%%%%%%%%%%%%%%%%%%%%%%%%%%%%%%%%%%%%%%%%%%%%%%%%
\begin{figure}[htb] \centering
\includegraphics[width=13cm,clip]{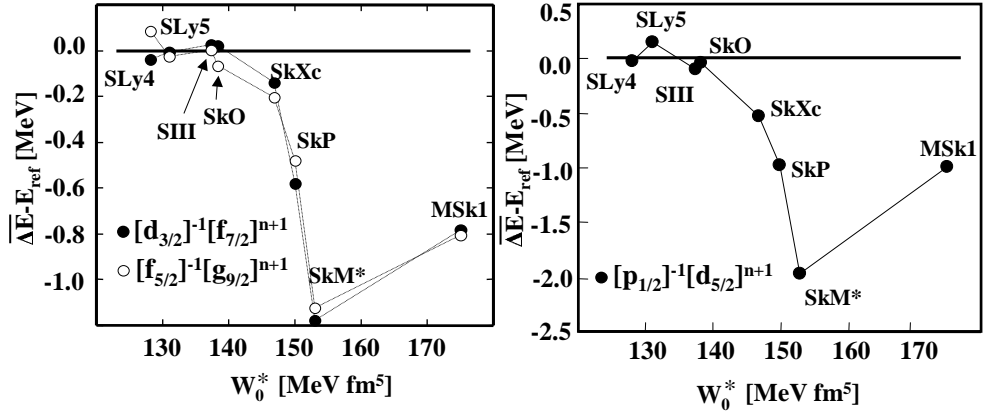}
\caption[]{Left figure shows mean value of
        $\overline{\Delta E} - E_{ref}$ in $A$$\sim$44 nuclei in
        comparison to analogical quantity calculated for
        $\Delta E = E([f_{5/2}^{-1}g_{9/2}^{n+1}])-E([g_{9/2}^{n}])$
        configurations in $A$$\sim$80 nuclei.
        Right figure shows
        $\overline{\Delta E} - E_{ref}$ calculated for
        $\Delta E = E([p_{1/2}^{-1}d_{5/2}^{n+1}])-E([d_{5/2}^{n}])$
        configurations in $A$$\sim$16 nuclei.
        See text for further details.}
\label{normal_ls}
\end{figure}
%%%%%%%%%%%%%%%%%%%%%%%%%%%%%%%%%%%%%%%%%%%%%%%%%%%%%%%%%%%%%%%%%%%%%%%%%%%%

%%%%%%%%%%%%%%%%%%%%%%%%%%%%%%%%%%%%%%%%%%%%%%%%%%%%%%%%%%%%%%%%%%%%%%%%%%%%
\begin{figure}[htb]
\centering
\includegraphics[width=8cm,clip]{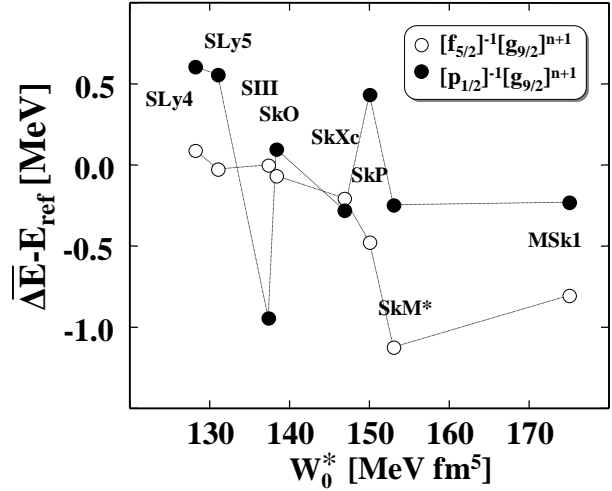}
\caption[]{Comparison of $\overline{\Delta E} - E_{ref}$ for two possible
particle-hole excitations $[f_{5/2}^{-1}g_{9/2}^{n+1}]$ and
$[p_{1/2}^{-1}g_{9/2}^{n+1}]$
through the magic gap 40.}
\label{A80b}
\end{figure}
%%%%%%%%%%%%%%%%%%%%%%%%%%%%%%%%%%%%%%%%%%%%%%%%%%%%%%%%%%%%%%%%%%%%%%%%%%%%%

To further investigate correlation between $\overline{\Delta E}$ and
the SO strength $W_0^*$ we have performed similar calculations
for terminating states $\Delta E =
E([p_{1/2}^{-1}d_{5/2}^{n+1}]_{I_{max}}) - E([d_{5/2}^{n}]_{I_{max}})$  in
$A$$\sim$16 mass region and for two different terminating configurations
$\Delta E = E([f_{5/2}^{-1}g_{9/2}^{n+1}]_{I_{max}}) -
E([g_{9/2}^{n}]_{I_{max}})$ and $\Delta E =
E([p_{1/2}^{-1}g_{9/2}^{n+1}]_{I_{max}}) - E([g_{9/2}^{n}]_{I_{max}})$ in
$A$$\sim$80 mass region. The results are depicted in Figs.~\ref{normal_ls}
and~\ref{A80b}. Since there is no data for terminating states in these nuclei
we compare purely theoretical trends by showing
mean value $\overline{\Delta E} - E_{ref}$ (averaging over different
$N$$\ne$$Z$ nuclei) where
$E_{ref}$ denotes mean value of $\overline{\Delta E}$ calculated using SkO,
SIII and Sly4 forces i.e. those which perform best in $A$$\sim$44 region.

Note, that dependence of $\overline{\Delta E} - E_{ref}$ on $W_0^*$
for $[p_{1/2}^{-1}d_{5/2}^{n+1}]$ and $[f_{5/2}^{-1}g_{9/2}^{n+1}]$
configurations
is very similar to each other and to $A$$\sim$44 case, as
shown in Fig.~\ref{normal_ls}. For
$[p_{1/2}^{-1}g_{9/2}^{n+1}]$ states one observes completely different
pattern, see~Fig.~\ref{A80b}. The reason for that becomes clear when
one looks at the hierarchy of different contributions to the corresponding
energy gaps emerging within Nilsson model:
\ben
        \label{ea16}
        \Delta e^{(d_{5/2}-p_{1/2})}_{N=Z=8} &=&
        \hbar \omega_0 (1 - 4\kappa ),
        \\
        \label{ea40}
        \Delta e^{(f_{7/2}-d_{3/2})}_{N=Z=20} &=&
        \hbar \omega_0 (1 - 6\kappa - 2 \kappa \mu),
        \\
        \label{ea80a}
        \Delta e^{(g_{9/2}-f_{5/2})}_{N=Z=40} &=&
        \hbar \omega_0 (1 - 8\kappa - 3 \kappa \mu),
        \\
        \label{ea80b}
        \Delta e^{(g_{9/2}-p_{1/2})}_{N=Z=40} &=&
        \hbar \omega_0 (1 - 6\kappa - 13 \kappa \mu).
\een
The value of $\mu$ vary from zero in light nuclei where the surface and flat
bottom effects are small to the pseudo-spin limit $\mu \sim 1/2$ in
heavy-nuclei. In (\ref{ea16}), (\ref{ea40}) and (\ref{ea80a}) contribution
to $\Delta e$ from flat bottom and surface
effect is small in comparison with contribution from SO interaction.
The ratio of these two terms (assuming $\mu = 1/2$) for (\ref{ea40})
and (\ref{ea80a}) cases is equal:
\be
        \frac{\mu}{3} \approx \frac{3 \mu}{8} \approx 0.17\, ,
        \nonumber
\ee
and for (\ref{ea16}) it is equal zero. In case of
\textit{ph} excitation from $p_{1/2}$ to $g_{9/2}$ sub-shell this ratio is:
\be
        \frac{13 \mu}{6} \approx 1,
\ee
and one may anticipate qualitative change in physical scenario what indeed
takes place as shown in Fig.~\ref{A80b}.

%----------------------------------------------------------------------------
%                                   Imax - 1
%----------------------------------------------------------------------------

\section{$I_{max}-1$ states}\label{i-1}

\begin{figure}[htb]
\centering
\includegraphics[width=13cm,clip]{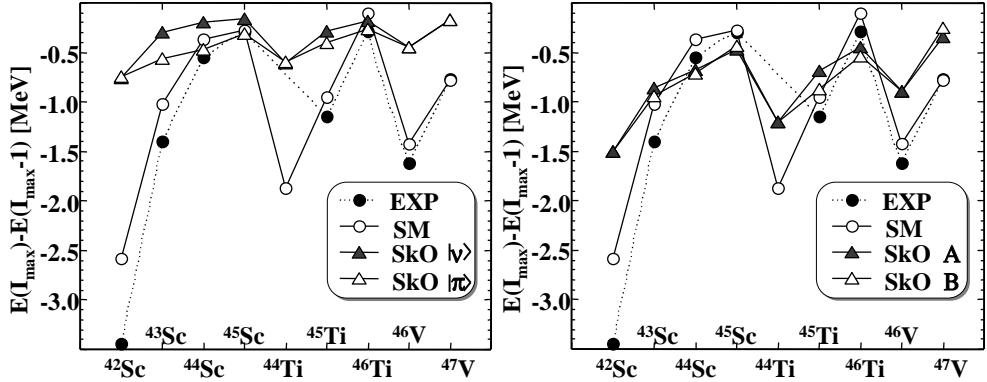}
\caption[]{Energy difference between $I_{max}$ and $I_{max}-1$ states
for $f_{7/2}^n$ configuration. Left panel shows
the SkO-SHF solutions $|\pi\rangle$ and $|\nu\rangle$ which break rotational
symmetry.  Right panel show the effect of symmetry restoration
according to the methods A and B, respectively.
In both cases results are compared with experimental data
(dots) and the SM calculations (circles).} \label{sko}
\end{figure}

Unfavored signature $[f_{7/2}^n]_{I_{max}-1}$ terminating states can be
obtained within mean-field theory by changing either the signature of
valence neutron or the signature of valence proton. These
two independent Slater determinants will be labeled
$|\nu\rangle$ and $|\pi\rangle$, respectively. Energy difference,
between these states and $I_{max}$ solution
calculated using modified SkO force is shown in the
left hand part of Fig.~\ref{sko}. Note that the SHF values are in
complete disagreement both with experimental data and SM results.
Similar result holds for SLy4 force, see Ref.~\cite{[Zal06]}.

The major source of disagreement is related to spontaneous
violation of rotational symmetry by these quasi-spherical SHF solutions.
Indeed,
simple $m$-scheme counting shows that both $I_{max}$ and
$I_{max}-1$ representations are single-folded within the $f_{7/2}^n$
Hilbert space irrespective of number of valence particles $n$.
The SHF solutions are  eigenstates of
an angular moment projection with $K=I_{max}-1$. Hence,
they are mixtures of "{\it spurious\/}" $|I_{max}, I_{max}-1\rangle$ and
{\it physical\/} $|I_{max}-1,I_{max}-1\rangle$ states:
\be\label{Psi_1}
        |I_{max},I_{max}-1\rangle = a |\nu\rangle +b |\pi\rangle ,
\ee
\be\label{Psi_2}
        |I_{max}-1,I_{max}-1\rangle = -b |\nu\rangle +a |\pi\rangle.
\ee
The symmetry restoration is therefore limited to a simple $2\times 2$
mixing problem:
\be\label{eigenequation}
        \left(\begin{array}{cc}
                e_1 & V \\
                V & e_2
        \end{array}\right)
        \left(\begin{array}{c}
                a \\ b
        \end{array}\right)
        = \lambda
        \left(\begin{array}{c}
                a \\ b
        \end{array}\right).
\ee
where $e_1\equiv E(|\nu\rangle)-E(I_{max})$ and $e_2\equiv
E(|\pi\rangle)-E(I_{max})$. The problem can be solved analytically in two
different ways. The first method (A) determines $V$ based on strict
requirement that ''{\it spurious\/}'' solution should be placed at
zero energy $\lambda_1\equiv 0$. This corresponds to $V=\sqrt{e_1e_2}$ and
$\lambda_2=e_1+e_2$. This is energy of physical $I_{max}-1$ state relative
to energy of $I_{max}$ state. The results of this procedure for the SkO-SHF
calculations are shown in the right hand side of Fig.~\ref{sko} (black
triangles).

%%%%%%%%%%%%%%%%%%%%%%%%%%%%%%%%%%%%%%%%%%%%%%%%%%%%%%%%%%%%%%%%%%%%%%%%%%%%
%%%%%%%%%%%%%%%%%%%%%%%%%%%%%%%%%%%%%%%%%%%%%%%%%%%%%%%%%%%%%%%%%%%%%%%%%%%%%

The method B uses values of $a$ and $b$ coefficients calculated
by acting on $|I_{max},I_{max}\rangle$ state with $\hat{I}_-$
operator, see Ref.~\cite{[Zal06]} for details. For example, in
$^{43}\mathrm{Sc}$ we obtain:
\be
\hat{I}_-|I_{max},I_{max}\rangle=|I_{max},I_{max}-1\rangle=
2\sqrt{3}|\pi\rangle+\sqrt{7}|\nu\rangle.
\ee
Knowing $a$ and $b$ one may find from (\ref{eigenequation}) the value
of interaction $V$ as well as energies of
''{\it spurious\/}'' and {\it physical\/} solutions
$\lambda_1$ and $\lambda_2$:
\ben
        V         &=& \frac{r(e_1-e_2)}{1-r^2}, \label{V2x2}     \\
        \lambda_1 &=& \frac{e_1 - r^2 e_2}{1-r^2}, \label{lamb1} \\
        \lambda_2 &=& \frac{e_2 - r^2 e_1}{1-r^2},  \label{lamb2}
\een
where $r \equiv b/a$.
Physical solution  $\lambda_2$ is shown the right hand side panel of
Fig.~\ref{sko}. Note, that in this case the energy, $\lambda_1$, of
''{\it spurious\/}''
state is not equal zero, but for SkO force it doesn't exceed $\pm
0.1~\mathrm{MeV}$ (with exception of $^{45}\mathrm{Ti}$).
Note also that method B doesn't work for $N=Z$
nuclei where $r=1$. It is
clearly seen from Fig~\ref{sko} that our simple symmetry
restoring schemes provide very accurate results. More
detailed discussion concerning this issue including results
for Sly4 force can be found in Ref.~\cite{[Zal06]}.

%%%%%%%%%%%%%%%%%%%%%%%%%%%%%%%%%%%%%%%%%%%%%%%%%%%%%%%%%%%%%%%%%%%%%%%%%%%%
\begin{figure}[htb]\label{imax-1}
\centering
\includegraphics[width=13cm,clip]{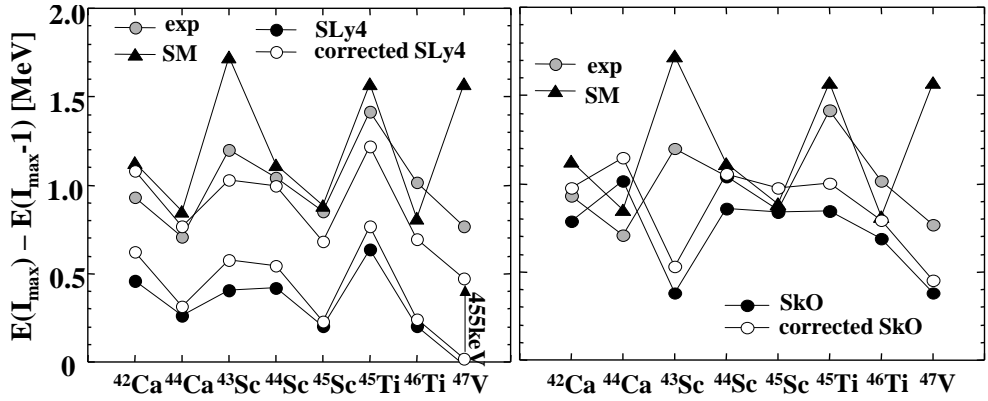}
\caption[]{Energy difference between
$I_{max}$ and the lowest $I_{max}-1$ terminating states for
$d_{3/2}^{-1}f_{7/2}^{n+1}$ configuration. Triangles and grey dots
denote SM results and experimental data respectively. Black dots represent
bare SHF solutions. Circles represent symmetry restored SHF results
according to the method A. Left (right) panel illustrates
the SLy4-SHF (SkO-SHF) results, respectively.
In order to facilitate comparison of isotopic and isotonic dependence
of the Sly4 results we include additional curve showing symmetry
restored SLy4-SHF solution shifted arbitrarily by 455\,keV.
The method B, which is not shown for the reasons of
clarity, gives larger correction than the method A by $\sim$100\,keV
for both Sly4 and SkO forces.} \end{figure}
%%%%%%%%%%%%%%%%%%%%%%%%%%%%%%%%%%%%%%%%%%%%%%%%%%%%%%%%%%%%%%%%%%%%%%%%%%%%%

For $d_{3/2}^{-1}f_{7/2}^{n+1}$ configuration
there are three possibilities of creating
$I_{max}-1$ states at the mean-field level. One may change the
signature of either proton or neutron in $f_{7/2}$ sub-shell.
These states will be
labeled as $|\pi\rangle$ and $|\nu\rangle$, respectively.
Alternatively, one may
change the signature of the proton in $d_{3/2}$
sub-shell. This Slater determinant will be denoted as
$|\bar{\pi}\rangle$.
Simple $M$-scheme counting method shows that our Hilbert space
contains two-fold $I_{max}-1$ representation.
Hence, to restore broken symmetry we have to deal here with
$3\times 3$ eigen-problem having two physical and one
''{\it spurious\/}'' $|I_{max},I_{max}-1\rangle$ solutions.
The most general mixing Hamiltonian matrix can in this case be
written as:
\be\label{Hcomplex}
        \left(
        \begin{array}{ccc}
                e_1      & V_{12}   & V_{13}^* \\
                V_{12}^* & e_2      & V_{23}   \\
                V_{13}   & V_{23}^* & e_2
        \end{array}
        \right)
        \left(
        \begin{array}{c}
                a \\
                b \\
                c
        \end{array}
        \right)=
        \lambda
        \left(
        \begin{array}{c}
                a \\
                b \\
                c
        \end{array}
        \right),
\ee
where $e_i \equiv E[i]-E[I_{max}]$, and $i=|\pi\rangle, |\nu\rangle,
|\bar{\pi}\rangle$. Two different
methods of restoring symmetry (called A and B) are proposed
below. In both methods the energy of ''{\it spurious\/}'' solution
is rigorously set to zero.
The method A assumes additionally real interaction and precise knowledge
of mixing coefficients $a$, $b$ and $c$. The values of $a$, $b$ and $c$
are calculated again by using simple angular momentum algebra,
in particular by applying $\hat{I}_-$ to
$|I_{max}, I_{max}\rangle$ reference state, see~\cite{[Zal06]} for
details. Method B admits complex interaction and sets no further
constraints on $a$, $b$ and $c$ coefficients.

In both cases the problem can be solved analytically. Method A gives:
\ben\label{int}
        V_{12} &=& \frac{-e_1a^2-e_2b^2+e_3c^2}{2ab}, \nonumber \\
        V_{13} &=& \frac{-e_1a^2+e_2b^2-e_3c^2}{2ac},  \\
        V_{23} &=& \frac{e_1a^2-e_2b^2-e_3c^2}{2bc}, \nonumber
\een
while in method B we obtain:
        \be
                V_{12}=\sqrt{e_1e_2},\quad
                V_{13}=\sqrt{e_1e_3},\quad
                V_{23}=\sqrt{e_2e_3},
        \ee
if $e_1, e_2, e_3$ are of the same sign or:
        \be
                V_{12}=i\sqrt{-e_1e_2},\quad
                V_{13}=-i\sqrt{-e_1e_3},\quad
                V_{23}=\sqrt{e_2e_3},
        \ee
for $e_1<0$ and $e_2, e_3>0$. Other possibilities are analogous.
Knowing the values of interaction one may find the energies of physical
$I_{max}-1$ states. They are equal:
\be
        \lambda_{\pm} = \frac{1}{2}
        \left(\Sigma \pm \sqrt{\Delta}\right)
\ee
where
\ben
        \Delta &=& \Sigma^2 - 4Z, \nonumber \\
        \Sigma &=& e_1+e_2+e_3, \nonumber \\
        Z &=& e_1e_2+e_1e_3+e_2e_3-|V_{12}|^2-|V_{13}|^2-|V_{23}|^2 \nonumber.
\een
The results for the lowest $I_{max}-1$ states are shown in
Fig.~\ref{imax-1}. One can see, that now the effect of
symmetry restoration is relatively small as compared to
clearly dominant mean-field splitting.
There is also an interesting difference
between Sly4 and SkO forces. For the Sly4
force the isotonic and isotopic dependence
of $E(I_{max})-E(I_{max}-1)$ is very well reproduced but there is a constant
offset between theoretical results and the data. This offset, after symmetry
restoration, equals to $\sim 455\,\mathrm{keV}$
for the method A and to $\sim 385\,\mathrm{keV}$ for the method B. The
opposite is true in the SkO case. This force does not reproduce details of
isotopic and isotonic dependence, but
reproduces quite well mean value of $E(I_{max})-E(I_{max}-1)$.

\section{Summary}

We have shown that terminating states in $A$$\sim$44 mass region
are excellent playground for testing and
constraining various aspects of the nuclear EDF. In particular, unification
of otherwise quite random coupling constants connected with spin-fields
to experimental data and slight reduction of
SO strength allow to reach very
good agreement with experimental data for maximum-spin $I_{max}$
terminating states. Our study reveals that the SHF method
cannot be directly applied to unfavored signature $I_{max}-1$
states. For these states the SHF solutions manifestly
break rotational invariance at almost spherical shape. After
identification of the underlying SSB mechanism we propose analytical
symmetry restoration schemes for both $[f_{7/2}^n]_{I_{max}-1}$ and
$[d_{3/2}^{-1} f_{7/2}^{n+1}]_{I_{max}-1}$ configurations.
It is shown that for $[f_{7/2}^n]_{I_{max}-1}$ the configuration
mixing is absolutely necessary in order to reproduce empirical
$E(I_{max})-E(I_{max}-1)$ splitting. For
$[d_{3/2}^{-1} f_{7/2}^{n+1}]_{I_{max}-1}$ configuration
on the other hand the splitting $E(I_{max})-E(I_{max}-1)$ is
dominated by mean-field itself and the symmetry restoration effect
is relatively less important.

\section{Acknowledgments}

This work was supported in part the Polish Committee for
Scientific Research (KBN) under contract No. 1~P03B~059~27 and  by the
Foundation for Polish Science (FNP).

%\bibliography{rev}
%\bibliographystyle{unsrt}

\end{document}